\documentclass[preprintnumbers,showpacs,amsmath,amssymb,prd,floatfix,twocolumn,superscriptaddress,nofootinbib]{revtex4}
\usepackage{graphicx}
\usepackage{epsfig}
\usepackage{bm}
\usepackage{amsfonts}

\def\rl{\rho_\Lambda}
\def\ol{\Omega_\Lambda}
\def\la{\Lambda}

\begin{document}

\title{Holographic dark energy with varying gravitational constant}

\author{Mubasher Jamil}
\email{mjamil@camp.edu.pk} \affiliation{Center for Advanced
Mathematics and Physics, National University of Sciences and
Technology,\\ Rawalpindi, 46000, Pakistan}

\author{Emmanuel N. Saridakis }
\email{msaridak@phys.uoa.gr} \affiliation{Department of Physics,
University of Athens, GR-15771 Athens, Greece}

\author{M. R. Setare}
\email{rezakord@ipm.ir} \affiliation{Department of Science, Payame
Noor University, Bijar, Iran }

\begin{abstract}
We investigate the holographic dark energy scenario with a varying
gravitational constant, in flat and non-flat background geometry.
We extract the exact differential equations determining the
evolution of the dark energy density-parameter, which include
$G$-variation correction terms. Performing a low-redshift
expansion of the dark energy equation of state, we provide the
involved parameters as functions of the current density
parameters, of the holographic dark energy constant and of the
$G$-variation.
\end{abstract}

 \maketitle

\section{Introduction}

Recent cosmological observations obtained by SNe Ia {\cite{c1}},
WMAP {\cite{c2}}, SDSS {\cite{c3}} and X-ray {\cite{c4}} indicate
that the universe experiences an accelerated expansion. Although
the simplest way to explain this behavior is the consideration of
a cosmological constant \cite{c7}, the two relevant problems
(namely the ``fine-tuning'' and the ``coincidence'' one) led to
the dark energy paradigm. The dynamical nature of dark energy, at
least in an effective level, can originate from various fields,
such is a canonical scalar field (quintessence) \cite{quint}, a
phantom field, that is a scalar field with a negative sign of the
kinetic term \cite{phant}, or the combination of quintessence and
phantom in a unified model named quintom \cite{quintom}.

Although going beyond the above effective description requires a
deeper understanding of the underlying theory of quantum gravity
\cite{Witten:2000zk} unknown at present, physicists can still make
some attempts to probe the nature of dark energy according to some
basic quantum gravitational principles. An example of such a
paradigm is the holographic dark energy scenario, constructed in
the light of the holographic principle
\cite{Cohen:1998zx,Horava:2000tb,Hsu:2004ri,Li:2004rb} (although
the recent developments in Horawa gravity could offer a dark
energy candidate with perhaps better quantum gravitational
foundations \cite{Horawa}). Its framework is the black hole
thermodynamics \cite{BH22} and the connection (known from AdS/CFT
correspondence) of the UV cut-of of a quantum field theory, which
gives rise to the vacuum energy, with the largest distance of the
theory \cite{Cohen:1998zx}. Thus, determining an appropriate
quantity $L$ to serve as an IR cut-off, imposing the constraint
that the total vacuum energy in the corresponding maximum volume
must not be greater than the mass of a black hole of the same
size, and saturating the inequality, one identifies the acquired
vacuum energy as holographic dark energy:
\begin{equation}\label{de}
\rho_\Lambda=\frac{3c^2}{8\pi G L^2},
\end{equation}
with $G$ the Newton's gravitational constant and $c$ a constant.
The holographic dark energy scenario has been tested and
constrained by various astronomical observations
\cite{obs3a,obs2,obs1,Wu:2007fs,obs3} and it has been extended to
various frameworks \cite{nonflat,holoext,intde}.

Until now, in all the investigated holographic dark energy models
a constant Newton's ``constant'' $G$ has been considered. However,
there are significant indications that $G$ can by varying, being a
function of time or equivalently of the scale factor \cite{G4com}.
In particular, observations of Hulse-Taylor binary pulsar
\cite{Damour,kogan}, helio-seismological data \cite{guenther},
Type Ia supernova observations \cite{c1}  and astereoseismological
data from the pulsating white dwarf star G117-B15A \cite{Biesiada}
lead to $\left|\dot{G}/G\right| \lessapprox 4.10 \times 10^{-11}
yr^{-1}$, for $z\lesssim3.5$ \cite{ray1}. Additionally, a varying
$G$ has some theoretical advantages too, alleviating the dark
matter problem \cite{gol}, the cosmic coincidence problem
\cite{jamil} and the discrepancies in Hubble parameter value
\cite{ber}.

There have been many proposals in the literature attempting to
theoretically justified a varying gravitational constant, despite
the lack of a full, underlying quantum gravity theory. Starting
with the simple but pioneering work of Dirac \cite{Dirac:1938mt},
the varying behavior in Kaluza-Klein theory was associated with a
scalar field appearing in the metric component corresponding to
the  $5$-th dimension \cite{kal} and its size variation
\cite{akk}. An alternative approach arises from Brans-Dicke
framework \cite{bd}, where the gravitational constant is replaced
by a scalar field coupling to gravity through a new parameter, and
it has been generalized to various forms of scalar-tensor theories
\cite{gen}, leading to a considerably broader range of
variable-$G$ theories. In addition, justification of a varying
Newton's constant has been established with the use of conformal
invariance and its induced local transformations \cite{bek}.
Finally, a varying $G$ can arise perturbatively through a
semiclassical treatment of Hilbert-Einstein action \cite{19},
non-perturbatively through quantum-gravitational approaches within
the ``Hilbert-Einstein truncation'' \cite{21}, or through
gravitational holography \cite{Guberina,7}.

In this work we are interested in investigating the holographic
dark energy paradigm allowing for a varying gravitational
constant, and extracting the corresponding corrections to the dark
energy equation-of-state parameter. In order to remain general and
explore the pure varying-$G$ effects in a model-independent way,
we do not use explicitly any additional, geometrical or
quintessence-like, scalar field, considering just the
Hilbert-Einstein action in an affective level, as it arises from
gravitational holography \cite{7,Guberina}. In other words, we
effectively focus on the dark energy and dark matter sectors
without examining explicitly the mechanism of G-variation, which
value is considered as an input fixed by observations.
Additionally, generality requires to perform our study in flat and
non-flat FRW universe. The plan of the work is as follows: In
section \ref{model} we construct the holographic dark energy
scenario with a varying Newton's constant and we extract the
differential equations that determine the evolution of dark energy
density-parameter. In section \ref{discussion} we use these
expressions in order to calculate the corrections to the dark
energy equation-of-state parameter at low redshifts. Finally, in
section \ref{Conclusions} we summarize our results.

\section{Holographic Dark Energy with varying gravitational constant}
\label{model}

\subsection{Flat FRW geometry}

Let us construct holographic dark energy scenario allowing for a
varying Newton's constant $G$. The space-time geometry will be a
flat Robertson-Walker:
\begin{equation}\label{met}
ds^{2}=-dt^{2}+a(t)^{2}(dr^{2}+r^{2}d\Omega^{2}),
\end{equation}
with $a(t)$ the scale factor and $t$ the comoving time. As usual,
the first Friedmann equation reads:
\begin{equation}\label{FR1}
H^2=\frac{8\pi G}{3}\Big(\rho_m+\rl\Big),
\end{equation}
with $H$ the Hubble parameter, $\rho_m=\frac{\rho_{m0}}{a^3}$,
where $\rho_m$ and $\rl$ stand respectively for  matter and dark
energy densities and the index $0$ marks the present value of a
quantity. Furthermore, we will use the density parameter $
\ol\equiv\frac{8\pi G}{3H^2}\rl$, which, imposing explicitly the
holographic nature of dark energy according to relation
(\ref{de}), becomes
\begin{eqnarray}
 \label{OmegaL2}
\ol=\frac{c^2}{H^2L^2}.
\end{eqnarray}
 Finally,
in the case of a flat universe, the best choice for the definition
of $L$ is to identify it with the future event horizon
\cite{Li:2004rb,Guberina,Guberina0,Hsu:2004ri}, that is $L\equiv
R_ h(a)$ with
\begin{equation}
 R_ h(a)=a\int_t^\infty{dt'\over
a(t')}=a\int_a^\infty{da'\over Ha'^2}~.\label{eh}
\end{equation}

In the following we will use $\ln a$ as an independent variable.
Thus, denoting by dot the time-derivative and by prime the
derivative with respect to $\ln a$, for every quantity $F$ we
acquire $\dot{F}=F'H$. Differentiating (\ref{OmegaL2}) using
(\ref{eh}), and observing that $\dot{R}_h=HR_h-1$, we obtain:
\begin{equation}\label{OmegaLdif}
\frac{\ol'}{\ol^2}=\frac{2}{\ol}\Big[-1-\frac{\dot{H}}{H^2}+\frac{\sqrt{\ol}}{c}\Big].
\end{equation}
Until now, the varying behavior of $G$ has not become manifested.
However, the next step is to eliminate $\dot{H}$. This can be
obtained by differentiating Friedman equation (\ref{FR1}), leading
to
\begin{equation}\label{Hdot}
2\frac{\dot{H}}{H^2}=-3+\ol\left(1+\frac{2\sqrt{\ol}}{c}\right)+\frac{G'}{G}\left(1-\ol\right),
\end{equation}
 where $G$ is considered to be a
function of $\ln a$. In the extraction of this relation we have
additionally used the auxiliary expression
\begin{equation}\label{rhoodot}
\rho'_\Lambda=\rho_\Lambda\left(-\frac{G'}{G}-2+\frac{2\sqrt{\ol}}{c}\right),
\end{equation}
which arises from differentiation of (\ref{de}).
 Therefore, substituting (\ref{Hdot}) into
(\ref{OmegaLdif}) we finally obtain:
\begin{equation}\label{OmegaLdif3}
\ol'=\ol(1-\ol)\Big[1+\frac{2\sqrt{\ol}}{c}\Big]-\ol(1-\ol)\frac{G'}{G}.
\end{equation}
The first term is the usual holographic dark energy differential
equation \cite{Li:2004rb}. The second term is the correction
arising from the varying nature of $G$. Note that $G'/G$ is a pure
number as expected.

Finally, for completeness, we present the general solution for
arbitrary $c$ and $G'/G\equiv\Delta_G $, which in an implicit form
 reads
\begin{eqnarray}
\label{solflat} \frac{\ln a}{c}
+x_0=\frac{\ln\Omega_\la}{c(1-\Delta_G)}
-\frac{\ln(1-\sqrt{\Omega_\la})}{2+c(1-\Delta_G)} +\ \ \ \ \ \ \ \
\ \nonumber\\ \ \
+\frac{\ln(1+\sqrt{\Omega_\la})}{2+c(\Delta_G-1)}
-\frac{8\ln[c(1-\Delta_G)+2\sqrt{\Omega_\la}]}{c(\Delta_G-1)[c^2(\Delta_G-1)^2-4]}.
\end{eqnarray}
The constant $x_0$ can be straightforwardly calculated if we
determine $a_0$ and $\Omega_{\la}^0$ today (for example choosing
$a_0=1$ $x_0$ is equal to the left hand side with $\Omega_\la$
replaced by $\Omega_\la^0$). Clearly, for $\Delta_G=0$ and $c=1$,
expression (\ref{solflat}) coincides with that of
\cite{Li:2004rb}.

\subsection{Non-flat FRW geometry}

In this subsection we generalize the aforementioned analysis in
the case of a general FRW universe with line element
\begin{equation}\label{metr}
 ds^{2}=-dt^{2}+a^{2}(t)\left(\frac{dr^2}{1-kr^2}+r^2d\Omega^{2}\right)
\end{equation}
in comoving coordinates  $(t,r,\theta,\varphi)$, where $k$ denotes
the spacial curvature with $k=-1,0,1$ corresponding to open, flat
and closed universe respectively. In this case, the first
Friedmann equation writes:
\begin{equation}\label{FR1nf}
H^2+\frac{k}{a^2}=\frac{8\pi G}{3}\Big(\rho_m+\rl\Big).
\end{equation}

According to the formulation of holographic dark energy in
non-flat geometry, the cosmological length $L$ in (\ref{OmegaL2})
is considered to be  \cite{nonflat}:
\begin{equation}\label{Lnonflat}
L\equiv\frac{a(t)}{\sqrt{|k|}}\,\text{sinn}\left(\frac{\sqrt{|k|}R_h}{a(t)}\right),
\end{equation}
where
\begin{equation}\frac{1}{\sqrt{|k|}}\text{sinn}(\sqrt{|k|}y)=
\begin{cases} \sin y  & \, \,k=+1,\\
             y & \, \,  k=0,\\
             \sinh y & \, \,k=-1.\\
\end{cases}\end{equation}
A straightforward calculation leads to
\begin{equation}\label{Ldot}
\dot{L}=H L-\text{cosn}\left(\frac{\sqrt{|k|}R_h}{a}\right),
\end{equation}
where
\begin{equation}\text{cosn}(\sqrt{|k|}y)=
\begin{cases} \cos y  & \, \,k=+1,\\
             1 & \, \,  k=0,\\
             \cosh y & \, \,k=-1.\\
\end{cases}\end{equation}

Repeating the procedure of the previous sub-section and
differentiating (\ref{OmegaL2}) using (\ref{Lnonflat}) and
(\ref{Ldot}) we obtain:
\begin{equation}\label{Hdotnf0}
\frac{\Omega^\prime_\Lambda}{\Omega^2_\Lambda}=\frac{2}{\Omega_\Lambda}
\left( -1-\frac{\dot H}{H^2}+\frac{\sqrt{\Omega_\Lambda}}{c}
\,\text{cosn}(\sqrt{|k|}y) \right).
\end{equation}
On the other hand, differentiating Friedmann equation
(\ref{FR1nf}) we finally obtain
\begin{eqnarray}\label{Hdotnf}
2\frac{\dot
H}{H^2}=-3-\Omega_k+\Omega_\Lambda\!\!&+&\!2\frac{\Omega_\Lambda^{3/2}}{c}\,
\text{cosn}\left(\frac{\sqrt{|k|}R_h}{a}\right)+\nonumber\\
&+&\!(1+\Omega_k-\ol)\frac{G^\prime}{G},
\end{eqnarray}
where we have introduced the curvature density parameter
$\Omega_k\equiv\frac{k}{(aH)^2}$. Therefore, substituting
(\ref{Hdotnf}) into (\ref{Hdotnf0}) we result to {\small{
\begin{eqnarray}
\label{Omprimenf}
\Omega_\Lambda^\prime=\Omega_\Lambda\!\left[\!1\!+\!\Omega_k\!-\!\Omega_\Lambda\!+\!\frac{2\sqrt{\Omega_\Lambda}}{c}\,
\text{cosn}\!\left(\frac{\sqrt{|k|}R_h}{a}\right)\!(1-\Omega_\Lambda)\right]\!-\nonumber\\
-\Omega_\Lambda(1+\Omega_k-\ol)\frac{G^\prime}{G}.\ \ \ \ \ \
\end{eqnarray}}}
Expression (\ref{Omprimenf}) provides the correction to
holographic dark energy differential equation in non-flat
universe, due to the varying nature of $G$. Clearly, for $k=0$
(and thus $\Omega_k=0$) it leads to (\ref{OmegaLdif3}).

\section{Cosmological implications}
\label{discussion}

Since we have extracted the expressions for $\ol'$, we can
calculate $w(z)$ for small redshifts $z$, performing the standard
expansions of the literature. In particular, since $\rho_\la\sim
a^{-3(1+w)}$ we acquire
\begin{equation}
\ln\rho_\la =\ln \rho_\la^0+{d\ln\rho_\la \over d\ln a} \ln a
+\frac{1}{2} {d^2\ln\rho_\la \over d(\ln a)^2}(\ln a)^2+\dots,
\end{equation}
 where the
derivatives are taken at the present time $a_0=1$ (and thus at
$\ol=\Omega_{\la}^0$). Then, $w(\ln a)$ is given as
\begin{equation}
w(\ln a)=-1-{1\over 3}\left[{d\ln\rho_\la \over d\ln a}
+\frac{1}{2} {d^2\ln\rho_\la \over d(\ln a)^2}\ln a\right],
\end{equation}
 up to second order.
 Since$\rho_\la =3H^2 \Omega_\la/(8\pi G)= \Omega_\la
\rho_m/\Omega_m=\rho_{m0}\Omega_\la/(1+\Omega_k -\Omega_\la)
a^{-3}$, the derivatives are easily computed using  the obtained
expressions for $\ol'$. In addition, we can straightforwardly
calculate $w(z)$, replacing $\ln a=-\ln(1+z)\simeq -z$, which is
valid for small redshifts, defining
\begin{equation}
w(z)=-1-{1\over 3}\left({d\ln\rho_\la \over d\ln a}\right)+
\frac{1}{6} \left[{d^2\ln\rho_\la \over d(\ln
a)^2}\right]\,z\equiv w_0+w_1z.
\end{equation}

The role of $G$-variation will be expressed through the pure
number $G'/G\equiv\Delta_G $, which will be extracted from
observations. In particular, observations of Hulse-Taylor binary
pulsar B$1913+16$ lead to the estimation
$\dot{G}/G\sim2\pm4\times10^{-12}{yr}^{-1}$ \cite{Damour,kogan},
while helio-seismological data provide the bound
$-1.6\times10^{-12}{yr}^{-1}<\dot{G}/G<0$ \cite{guenther}.
Similarly,  Type Ia supernova observations \cite{c1}  give the
best upper bound of the variation of $G$ as $-10^{-11} yr^{-1}
\leq \frac{\dot G}{G}<0$ at redshifts $z \simeq 0.5$
\cite{Gaztanaga}, while astereoseismological data from the
pulsating white dwarf star G117-B15A lead to $\left|\frac{\dot
G}{G}\right| \leq 4.10 \times 10^{-11} yr^{-1}$ \cite{Biesiada}.
See also \cite{ray1} for various bounds on $\dot{G}/G$ from
observational data, noting that all these measurements are valid
at relatively low redshifts, i.e $z\lesssim3.5$.

Since the limits in $G$-variation are given for $\dot{G}/G$ in
units $yr^{-1}$, and since $\dot{G}/G=H G'/G$, we can estimate
$\Delta_G $ substituting the value of $H$ in $yr^{-1}$. In the
following we will use $\left|\frac{\dot G}{G}\right| \lesssim 4.10
\times 10^{-11} yr^{-1}$. Thus, inserting an average estimation
for the Hubble parameter $H\approx\langle H\rangle\approx 6 \times
10^{-11} yr^{-1}$ \cite{Zhang:2009ae}, we obtain that
$0<|\Delta_G|\lesssim0.07$. Clearly, this estimation is valid at
low redshifts, since only in this range the measurements of $
\dot{G}/G$ and the estimation of the average $\langle H\rangle$
are valid. However, the restriction to this range is consistent
with the $z$-expansion of $w$ considered above.

\subsection{Flat FRW geometry}

In this case $\ol'$ is given by (\ref{OmegaLdif3}), and the
aforementioned procedure leads to
\begin{eqnarray}
&&w_0=-{1\over 3}-{2\over 3c}\sqrt{\Omega_\la^0}
+\frac{\Delta_G}{3}\label{w0fl}\\
\label{w1fl}
 &&w_1={1\over
6c}\sqrt{\Omega_\la^0}(1-\Omega_\la^0)\left(1+{2\over
c}\sqrt{\Omega_\la^0}\right)-\nonumber\\
&&\ \ \ \ \ \ \ \ \ \ \ \ \ \   \ \ \ \ \ \ \ \ \ \ \ \
-\frac{(1-\Omega_\la^0)\sqrt{\Omega_\la^0}}{6c}\Delta_G. \ \ \ \ \
\ \   \ \ \ \ \
\end{eqnarray}
These expressions provide $w_0$ and $w_1$, for the holographic
dark energy with varying $G$, in a flat universe. Obviously, when
$\Delta_G=0$, they coincide with those of \cite{Li:2004rb}.

In general, apart from the relevant uncertainty in $\Omega_\la^0$
measurements, we face the problem of the uncertainty in the
constant $c$. In particular, observational data from type Ia
supernovae give the best-fit value $c=0.21$  within 1-$\sigma$
error range \cite{obs3a}, while those from the X-ray gas mass
fraction of galaxy clusters lead to $c=0.61$ within 1-$\sigma$
\cite{obs2}. Similarly, combining data from type Ia supernovae,
Cosmic Microwave Background radiation and large scale structure
give the best-fit value $c=0.91$ within 1-$\sigma$ \cite{obs1},
while combining data from type Ia supernovae, X-ray gas and Baryon
Acoustic Oscillation lead to $c=0.73$ as a best-fit value within
1-$\sigma$ \cite{Wu:2007fs}. However, expressions
(\ref{w0fl}),(\ref{w1fl}) provide the pure change due to the
variation of gravitational constant for given $c$ and $\ol^0$. For
example, and in order to compare with the corresponding result of
\cite{Li:2004rb}, imposing $\Omega_\la^0\approx0.73$ and $c=1$,
and using $0<|\Delta_G|<0.07$ we obtain:
\begin{eqnarray}
&&w_0=-0.903^{+0.023}_{-0.023}\nonumber\\
&&w_1=0.1041^{+0.0025}_{-0.0025},
\end{eqnarray}
where we have neglected uncertainties other than $G$-variation.
Finally, note that the $w_0$-variation due to $\Delta_G$ is
absolute, that is it does not depend on $c$ and $\Omega_\la^0$,
while that of $w_1$ does depend on these parameters. However, the
relative variations of $w_0$,$w_1$ do depend on the $c$-value, and
they are smaller for smaller $c$.

\subsection{Non-flat FRW geometry}

In this case $\ol'$ is given by (\ref{Omprimenf}), and the
aforementioned procedure leads to
\begin{widetext}
\begin{eqnarray}\label{w0nonfl}
&&w_0=-{1\over 3}-{2\over
3c}\sqrt{\Omega_\la^0}\,\text{cosn}\frac{\sqrt{|k|}R_{h0}}{a_0}
+\frac{\Delta_G}{3}\\
\label{w1nonfl}
&&w_1=\frac{\sqrt{\Omega_\la^0}}{6c}\left[1+\Omega_k^0-\Omega_\Lambda^0+\frac{2\sqrt{\Omega_\Lambda^0}}{c}\,
\text{cosn}\left(\frac{\sqrt{|k|}R_{h0}}{a_0}\right)(1-\Omega_\Lambda^0)\right]\text{cosn}\left(\frac{\sqrt{|k|}R_{h0}}{a_0}\right)+
\frac{\Omega_\la^0}{3c^2}\,q\left(\frac{\sqrt{|k|}R_{h0}}{a_0}\right)-\nonumber\\
&&\ \ \ \ \ \ \ \   \ \ \ \ \ \ \ \ \ \ \ \ \ \ \ \ \ \ \   \ \ \
\ \ \ \ \ \ \ \ \ \ \ \ \ \ \ \   \ \ \ \ \ \ \ \ \ \ \ \ \ \ \ \
\ \ \   \ \ \ \ \ \ \ \ \ \ \ \ \ \
-\frac{\sqrt{\Omega_\la^0}}{6c}(1+\Omega_k^0-\ol^0)\text{cosn}\left(\frac{\sqrt{|k|}R_{h0}}{a_0}\right)\Delta_G.
\end{eqnarray}
\end{widetext}
In these expressions, $\Omega_k^0$ is the present day value of the
curvature density parameter, and we have defined
\begin{equation}
q(\sqrt{|k|}y)=
\begin{cases} \sin^2 y  & \, \,k=+1,\\
             0 & \, \,  k=0,\\
             -\sinh^2 y & \, \,k=-1.\\
\end{cases}\end{equation}
Finally, $R_{h0}$ and $a_0$ are the present values of the
corresponding quantities. Clearly, for $k=0$, that is for a flat
geometry, (\ref{w0nonfl}),(\ref{w1nonfl}) coincide with
(\ref{w0fl}),(\ref{w1fl}) respectively.

As we observe, expressions (\ref{w0nonfl}),(\ref{w1nonfl}), apart
from the present values of the parameters  $\Omega_\Lambda^0$,
$\Omega_k^0$ contain  $a_0$ and the value of $R_{h0}$ at present.
This last term is present in a non-flat universe, and it is a
``non-local'' quantity which has to be calculated by an
integration (see relations (\ref{Lnonflat}) and (\ref{eh})).
However, making use of the holographic nature of dark energy, we
can overcome this difficulty. Indeed, from (\ref{OmegaL2}) we
obtain that $L_0=c/(H_0\sqrt{\Omega_\Lambda^0})$, with $H_0$ the
present value of the Hubble parameter. On the other hand, from
(\ref{Lnonflat}) we acquire
$R_{h0}/a_0=\frac{1}{\sqrt{|k|}}\text{sinn}^{-1}(\sqrt{|k|}L_0/a_0)$.
Therefore, we conclude that
\begin{eqnarray}
\frac{R_{h0}}{a_0}&=&\frac{1}{\sqrt{|k|}}\text{sinn}^{-1}\left(\frac{c\sqrt{|k|}}{a_0H_0\sqrt{\Omega_\Lambda^0}}\right)=
\nonumber\\
&=&
\frac{1}{\sqrt{|k|}}\text{sinn}^{-1}\left(\frac{c\sqrt{|\Omega_k^0|}}{\sqrt{\Omega_\Lambda^0}}\right),
\end{eqnarray}
a relation which proves very useful. Substituting  into
(\ref{w0nonfl}),(\ref{w1nonfl}) we finally obtain the simple
expressions:
\begin{widetext}
\begin{eqnarray}\label{w0nonflb}
&&w_0=-{1\over 3}-{2\over 3c}\sqrt{\Omega_\la^0-c^2\Omega_k^0}
+\frac{\Delta_G}{3}\\
\label{w1nonflb}
&&w_1=\frac{\Omega_k^0}{3}+\frac{1}{6c}\sqrt{\Omega_\la^0-c^2\Omega_k^0}
\left[1+\Omega_k^0-\Omega_\Lambda^0+\frac{2}{c}(1-\ol)\sqrt{\Omega_\la^0-c^2\Omega_k^0}\right]-
\frac{1}{6c}\sqrt{\Omega_\la^0-c^2\Omega_k^0}
\left(1+\Omega_k^0-\Omega_\Lambda^0\right)\Delta_G.\ \
\end{eqnarray}
\end{widetext}
Note that $w_0$,$w_1$ depend eventually only on
$\Omega_\Lambda^0$, $\Omega_k^0$, $c$ and of course $\Delta_G$.
Similarly to the previous subsection, in order to give a
representative estimation and neglecting uncertainties of other
quantities apart from $G$-variation, we use $c=1$,
$\Omega_\la^0\approx0.73$, $\Omega_k^0\approx0.02$,
$0<|\Delta_G|<0.07$, obtaining:
\begin{eqnarray}
&&w_0=-0.895^{+0.023}_{-0.023}\nonumber\\
&&w_1=0.111^{+0.003}_{-0.003}.
\end{eqnarray}
Finally, we mention that the relative variations of $w_0$,$w_1$
depend on the $c$-value, and they are smaller for smaller $c$.

\section{Conclusions}
\label{Conclusions}

In this work we have investigated the holographic dark energy
scenario with a varying gravitational constant, going beyond the
simple scenarios of \cite{Horvat}. Imposing flat and non-flat
background geometry we have extracted the exact differential
equations that determine the evolution of the dark energy
density-parameter, where the $G$-variation appears as a
coefficient in additional terms. Thus, performing a low-redshift
expansion of the dark energy equation-of-state parameter
$w(z)\approx w_0 +w_1 z$, we provide $w_0$,$w_1$ as functions of
$\Omega_\la^0$, $\Omega_k^0$, of the holographic dark energy
constant $c$, and of the $G$-variation $\Delta_G$ (expressions
(\ref{w0nonflb}),(\ref{w1nonflb})). As expected, the variation of
the gravitational constant increases the variation of $w(z)$.

In the aforementioned analysis, the $G$-variation has been
considered as a constant quantity at the cosmological epoch of
interest, that is at low redshifts, as it is measured in
observations with satisfactory accuracy
\cite{Damour,kogan,guenther,Gaztanaga,Biesiada,ray1}. A step
forward would be to consider possible $G(z)$-parametrizations
\cite{Galli:2009pr,Xu:2009ss} and extract their effect on $w(z)$.
However, such parametrizations have a significant amount of
arbitrariness, since the present observational data do not allow
for such a resolution, and thus we have not performed this
extension in the present work.

Finally, we mention that in general, the possible uncertainty of
the constant $c$ can have a larger effect on $w(z)$ than that of
$G$-variation. In the above investigation we have just provided
the complete expressions, including the correction terms due to
the variation of the gravitational constant. One could proceed to
a combined observational constraint analysis, allowing for
variations and uncertainties in all parameters, as it was
partially performed in the specific Brans-Dicke framework in
\cite{Galli:2009pr}. This extended examination, with
not-guaranteed results due to complexity, is under current
investigation and it is left for a future publication.


\begin{thebibliography}{0}

\bibitem{c1}
 A.~G.~Riess {\it et al.}  [Supernova Search Team Collaboration],
  Astron.\ J.\  {\bf 116}, 1009 (1998);
S. Perlmutter {\it et al.} [Supernova Cosmology Project
Collaboration], Astrophys. J. {\bf 517}, 565 (1999).

\bibitem{c2}
C. L. Bennett {\it et al.}, Astrophys. J. Suppl. {\bf 148}, 1
(2003).
\bibitem{c3}
M. Tegmark {\it et al.} [SDSS Collaboration], Phys. Rev. D {\bf
69}, 103501 (2004).
\bibitem{c4}
S. W. Allen, \emph{et al.}, Mon. Not. Roy. Astron. Soc. {\bf 353},
457 (2004).



\bibitem{c7}
V. Sahni and A. Starobinsky, Int. J. Mod. Phy. D {\bf 9}, 373
(2000); P. J. Peebles and B. Ratra, Rev. Mod. Phys. {\bf 75}, 559
(2003).


\bibitem{quint}
B.~Ratra and P.~J.~E.~Peebles, Phys.\ Rev.\ D {\bf 37}, 3406
(1988); C.~Wetterich, Nucl.\ Phys.\ B {\bf 302}, 668 (1988);
A.~R.~Liddle and R.~J.~Scherrer, Phys.\ Rev.\ D {\bf 59}, 023509
(1999); I.~Zlatev, L.~M.~Wang and P.~J.~Steinhardt, Phys.\ Rev.\
Lett.\ {\bf 82}, 896 (1999); Z.~K.~Guo, N.~Ohta and Y.~Z.~Zhang,
Mod.\ Phys.\ Lett.\  A {\bf 22}, 883 (2007);
%\cite{Dutta:2009yb}
  S.~Dutta, E.~N.~Saridakis and R.~J.~Scherrer,
Phys.\ Rev.\  D {\bf 79}, 103005 (2009) [arXiv:0903.3412
[astro-ph.CO]].
  %%CITATION = ARXIV:0903.3412;%%

\bibitem{phant} R. R. Caldwell, Phys.
Lett. B {\bf{545}}, 23 (2002); R.~R.~Caldwell, M.~Kamionkowski and
N.~N.~Weinberg, Phys. Rev. Lett. {\bf 91}, 071301 (2003); S.
Nojiri and S. D. Odintsov, Phys. Lett. B {\bf 562}, 147 (2003); V.
K. Onemli and R. P. Woodard, Phys.\ Rev.\ D {\bf 70}, 107301
(2004); M. R. Setare, J. Sadeghi, A. R. Amani, Phys. Lett. B {\bf
666}, 288, (2008);
%\cite{Chen:2008ft}
  X.~m.~Chen, Y.~g.~Gong and E.~N.~Saridakis,
  %``Phase-space analysis of interacting phantom cosmology,''
  JCAP {\bf 0904}, 001 (2009);
  %%CITATION = JCAPA,0904,001;%%
%\cite{Saridakis:2009pj}
  E.~N.~Saridakis,
  %``Phantom evolution in power-law potentials,''
  Nucl.\ Phys.\  B {\bf 819}, 116 (2009).
  %%CITATION = NUPHA,B819,116;%%


\bibitem{quintom}
B.~Feng, X.~L.~Wang and X.~M.~Zhang, Phys.\ Lett.\  B {\bf 607}, 35
(2005);
  %%CITATION = PHLTA,B607,35;%%
Z. K. Guo, {\it{et al.}}, Phys. Lett. B {\bf 608}, 177 (2005); M.-Z
Li, B. Feng, X.-M Zhang, JCAP, 0512, 002 (2005); B. Feng, M. Li,
Y.-S. Piao and X. Zhang, Phys. Lett. B {\bf 634}, 101 (2006); M. R.
Setare, Phys. Lett. B {\bf 641}, 130 (2006); W. Zhao and Y. Zhang,
Phys. Rev. D {\bf73}, 123509 (2006);
 M. R.
Setare, J. Sadeghi, and A. R. Amani, Phys. Lett. B {\bf 660}, 299
(2008); J. Sadeghi, M. R. Setare, A. Banijamali and F. Milani, Phys.
Lett. B {\bf 662}, 92 (2008);
%\cite{Setare:2008pz}
  M.~R.~Setare and E.~N.~Saridakis,
  Phys.\ Lett.\  B {\bf 668}, 177 (2008);
  %%CITATION = PHLTA,B668,177;%%
 %\cite{Setare:2008si}
  M.~R.~Setare and E.~N.~Saridakis,
  JCAP {\bf 0809}, 026 (2008);
  %\cite{Setare:2008dw}
  M.~R.~Setare and E.~N.~Saridakis,
  %``Quintom Cosmology with General Potentials,''
  Int.\ J.\ Mod.\ Phys.\  D {\bf 18}, 549 (2009).
  %%CITATION = IMPAE,D18,549;%

\bibitem{Witten:2000zk}
  E.~Witten,
  %``The cosmological constant from the viewpoint of string theory,''
  hep-ph/0002297.
  %%CITATION = HEP-PH 0002297;%%


%\cite{Cohen:1998zx}
\bibitem{Cohen:1998zx}
  A.~G.~Cohen, D.~B.~Kaplan and A.~E.~Nelson,
  %``Effective field theory, black holes, and the cosmological constant,''
  Phys.\ Rev.\ Lett.\  {\bf 82}, 4971 (1999).
  %%CITATION = HEP-TH 9803132;%%


\bibitem{Horava:2000tb}
  G.~'t Hooft,
  %``Dimensional reduction in quantum gravity,''
  gr-qc/9310026;
  %%CITATION = GR-QC 9310026;%%
  L.~Susskind,
  %``The World as a hologram,''
  J.\ Math.\ Phys.\  {\bf 36}, 6377 (1995);
  P.~Horava and D.~Minic,
  %``Probable values of the cosmological constant in a holographic theory,''
  Phys.\ Rev.\ Lett.\  {\bf 85}, 1610 (2000);
  %%CITATION = HEP-TH 0001145;%%
  S.~D.~Thomas,
  %``Holography stabilizes the vacuum energy,''
  Phys.\ Rev.\ Lett.\  {\bf 89}, 081301 (2002).
  %%CITATION = PRLTA,89,081301;%%




%\cite{Hsu:2004ri}
\bibitem{Hsu:2004ri}
  S.~D.~H.~Hsu,
  %``Entropy bounds and dark energy,''
  Phys.\ Lett.\ B {\bf 594}, 13 (2004).
  %%CITATION = HEP-TH 0403052;%%

%\cite{Li:2004rb}
\bibitem{Li:2004rb}
  M.~Li,
  %``A model of holographic dark energy,''
  Phys.\ Lett.\ B {\bf 603}, 1 (2004).
  %%CITATION = HEP-TH 0403127;%%

\bibitem{Horawa}
  P.~Horava,
  Phys.\ Rev.\  D {\bf 79}, 084008 (2009);
  G.~Calcagni,
  arXiv:0904.0829 [hep-th];
  H.~Lu, J.~Mei and C.~N.~Pope,
  arXiv:0904.1595 [hep-th];
   C.~Charmousis, G.~Niz, A.~Padilla and P.~M.~Saffin,
  arXiv:0905.2579 [hep-th];
%\cite{Saridakis:2009bv}
  E.~N.~Saridakis,
  arXiv:0905.3532 [hep-th];
  %%CITATION = ARXIV:0905.3532;%%
  X.~Gao, Y.~Wang, R.~Brandenberger and A.~Riotto,
  arXiv:0905.3821 [hep-th];
  M.~i.~Park,
  arXiv:0905.4480 [hep-th];
  Y.~F.~Cai and E.~N.~Saridakis,
  arXiv:0906.1789 [hep-th];
  %%CITATION = ARXIV:0906.1789;%%
  M.~Botta-Cantcheff, N.~Grandi and M.~Sturla,
  arXiv:0906.0582 [hep-th].

\bibitem{BH22}
R.~C.~Myers and M.~J.~Perry, Annals Phys.  {\bf 172}, 304 (1986);
 P.~Kanti and
K.~Tamvakis, Phys. Rev. D {\bf 68}, 024014 (2003).

\bibitem{obs3a}
  Q.~G.~Huang and Y.~G.~Gong,
  %``Supernova constraints on a holographic dark energy model,''
  JCAP {\bf 0408}, 006 (2004);

\bibitem{obs2}
  Z.~Chang, F.~Q.~Wu and X.~Zhang,
  %``Constraints on holographic dark energy from X-ray gas mass fraction of
  %galaxy clusters,''
  Phys.\ Lett.\ B {\bf 633}, 14 (2006).
  %%CITATION = ASTRO-PH 0509531;%%

\bibitem{obs1}
  X.~Zhang and F.~Q.~Wu,
  %``Constraints on holographic dark energy from Type Ia supernova
  %observations,''
  Phys.\ Rev.\ D {\bf 72}, 043524 (2005).
  %%CITATION = ASTRO-PH 0506310;%%


%\cite{Wu:2007fs}
\bibitem{Wu:2007fs}
  Q.~Wu, Y.~Gong, A.~Wang and J.~S.~Alcaniz,
  Phys.\ Lett.\  B {\bf 659}, 34 (2008);
  Y.~Z.~Ma and Y.~Gong,
  Eur.\ Phys.\ J.\  C {\bf 60}, 303 (2009).
  %%CITATION = EPHJA,C60,303;%%

\bibitem{obs3}
  K.~Enqvist, S.~Hannestad and M.~S.~Sloth,
  %``Searching for a holographic connection between dark energy and the  low-l
  %CMB multipoles,''
  JCAP {\bf 0502} 004 (2005);
  %%CITATION = ASTRO-PH 0409275;%%
  J.~Shen, B.~Wang, E.~Abdalla and R.~K.~Su,
  %``Constraints on the dark energy from the holographic connection to the
  %small l CMB suppression,''
  Phys.\ Lett.\ B {\bf 609} 200 (2005);
  %%CITATION = HEP-TH 0412227;%%
  H.~C.~Kao, W.~L.~Lee and F.~L.~Lin
  %``CMB constraints on the holographic dark energy model,''
  Phys.\ Rev.\ D {\bf 71} 123518 (2005).
  %%CITATION = ASTRO-PH 0501487;%%


%\cite{nonflat}
\bibitem{nonflat}
 Q.~G.~Huang and M.~Li,
 % ``The holographic dark energy in a non-flat universe,''
  JCAP {\bf 0408}, 013 (2004).
  %%CITATION = ASTRO-PH 0404229;%%



%\cite{holoext}
\bibitem{holoext}
 M.~Ito,
  %``Holographic dark energy model with non-minimal coupling,''
 Europhys.\ Lett.\  {\bf 71}, 712 (2005);
  %%CITATION = HEP-TH 0405281;%%
  K.~Enqvist and M.~S.~Sloth,
  %``A CMB / dark energy cosmic duality,''
  Phys.\ Rev.\ Lett.\  {\bf 93}, 221302 (2004);
  %%CITATION = HEP-TH 0407056;%%
  Q.~G.~Huang and M.~Li,
  %``Anthropic principle favors the holographic dark energy,''
  JCAP {\bf 0503}, 001 (2005);
  %%CITATION = HEP-TH 0410095;%%
  D.~Pavon and W.~Zimdahl,
  %``Holographic dark energy and cosmic coincidence,''
  Phys.\ Lett.\ B {\bf 628}, 206 (2005);
  %%CITATION = GR-QC 0505020;%%
  B.~Wang, Y.~Gong and E.~Abdalla,
  %``Transition of the dark energy equation of state in an interacting
  %holographic dark energy model,''
  Phys.\ Lett.\ B {\bf 624}, 141 (2005);
  %%CITATION = HEP-TH 0506069;%%
  H.~Kim, H.~W.~Lee and Y.~S.~Myung,
  %``Equation of state for an interacting holographic dark energy model,''
  Phys.\ Lett.\ B {\bf 632}, 605 (2006);
  %%CITATION = GR-QC 0509040;%%
  S.~Nojiri and S.~D.~Odintsov,
  % ``Unifying phantom inflation with late-time acceleration: Scalar
  % phantom-non-phantom transition model and generalized holographic dark
  %energy,''
  Gen.\ Rel.\ Grav.\  {\bf 38}, 1285 (2006);
  %%CITATION = HEP-TH 0506212;%%
  E.~Elizalde, S.~Nojiri, S.~D.~Odintsov and P.~Wang,
   %``Dark energy: Vacuum fluctuations, the effective phantom phase, and
  %holography,''
  Phys.\ Rev.\ D {\bf 71}, 103504 (2005);
  %%CITATION = HEP-TH 0502082;%%
  B.~Hu and Y.~Ling,
  %``Interacting dark energy, holographic principle and coincidence problem,''
  Phys.\ Rev.\ D {\bf 73}, 123510 (2006);
  %%CITATION = HEP-TH 0601093;%%
  H.~Li, Z.~K.~Guo and Y.~Z.~Zhang,
  %``A tracker solution for a holographic dark energy model,''
  Int.\ J.\ Mod.\ Phys.\ D {\bf 15}, 869 (2006);
  %%CITATION = GR-QC 0606103;%%
  M.~R.~Setare,
  %``Interacting holographic dark energy model in non-flat universe,''
  Phys.\ Lett.\ B {\bf 642}, 1 (2006);
  %%CITATION = HEP-TH 0609069;%%
  M.~R.~Setare, Phys. Lett. {\bf B642}, 421, (2006);
%\cite{Saridakis:2007cy}
  E.~N.~Saridakis,
  %``Restoring Holographic Dark Energy in Brane Cosmology,''
  Phys.\ Lett.\  B {\bf 660}, 138 (2008);
  %%CITATION = PHLTA,B660,138;%%
%\cite{Saridakis:2007ns}
  E.~N.~Saridakis,
  %``Holographic Dark Energy in Braneworld Models with Moving Branes and the
  %w=-1 Crossing,''
  JCAP {\bf 0804}, 020 (2008);
  %%CITATION = JCAPA,0804,020;%%
%\cite{Saridakis:2007wx}
  E.~N.~Saridakis,
  %``Holographic Dark Energy in Braneworld Models with a Gauss-Bonnet Term in
  %the Bulk. Interacting Behavior and the w =-1 Crossing,''
  Phys.\ Lett.\  B {\bf 661}, 335 (2008).
  %%CITATION = PHLTA,B661,335;%%

  \bibitem{intde}
  L.~Amendola,
  %``Coupled quintessence,''
  Phys.\ Rev.\ D {\bf 62}, 043511 (2000);
  D.~Comelli, M.~Pietroni and A.~Riotto,
  %``Dark energy and dark matter,''
  Phys.\ Lett.\ B {\bf 571}, 115 (2003);
   M. R. Setare, JCAP 0701, 023, (2007);
    M. Jamil and M.A. Rashid,
Eur. Phys. J. C 60 (2009) 141; ibid, Eur.Phys.J.C58:111-114,2008;
%\cite{Setare:2008hm}
  M.~R.~Setare and E.~N.~Saridakis,
  %``Correspondence between Holographic and Gauss-Bonnet dark energy models,''
  Phys.\ Lett.\  B {\bf 670}, 1 (2008).
  %%CITATION = PHLTA,B670,1;%%

\bibitem{G4com}
S.~D'Innocenti, G.~Fiorentini, G.~G.~Raffelt, B.~Ricci and
A.~Weiss, Astron. Astrophys.  {\bf 312}, 345 (1996); K.~Umezu,
K.~Ichiki and M.~Yahiro, Phys. Rev. D {\bf 72}, 044010 (2005);
 S.~Nesseris and L.~Perivolaropoulos, Phys. Rev. D {\bf 73}, 103511
(2006); J.~P.~W.~Verbiest {\it et al.} [arXiv:astro-ph/0801.2589].

\bibitem{kogan}
  G.~S.~Bisnovatyi-Kogan,
  %``Checking the variability of the gravitational constant with binary
  %pulsars,''
  Int.\ J.\ Mod.\ Phys.\  D {\bf 15}, 1047 (2006).
  %%CITATION = IMPAE,D15,1047;%%


\bibitem{Damour}
Damour T.,{\it  et al}, Phys. Rev. Lett. {\bf 61}, 1151 (1988).

\bibitem{guenther} D.B. Guenther, Phys. Lett. B {\bf 498}, 871 (1998).

\bibitem{Gaztanaga}
  E.~Gaztanaga, E.~Garcia-Berro, J.~Isern, E.~Bravo and I.~Dominguez,
  Phys.\ Rev.\  D {\bf 65}, 023506 (2002).
  %%CITATION = PHRVA,D65,023506;%%


\bibitem{Biesiada} Biesiada M. and Malec B.,  Mon. Not. R. Astron. Soc. {\bf 350}, 644
(2004).



\bibitem{ray1}
  S.~Ray and U.~Mukhopadhyay,
  %``Dark energy models with time-dependent gravitational constant,''
  Int.\ J.\ Mod.\ Phys.\  D {\bf 16}, 1791 (2007).


\bibitem{gol}I. Goldman, Phys. Lett. B {\bf281}, 219 (1992).

\bibitem{jamil}
  M.~Jamil, F.~Rahaman and M.~Kalam,
  Eur.\ Phys.\ J.\  C {\bf 60}, 149 (2009).
  %%CITATION = EPHJA,C60,149;%%



\bibitem{ber}O. Bertolami et al, Phys. Lett. B {\bf311}, 27 (1993).

%\cite{Dirac:1938mt}
\bibitem{Dirac:1938mt}
  P.~A.~M.~Dirac,
  %``New basis for cosmology,''
  Proc.\ Roy.\ Soc.\ Lond.\  A {\bf 165} (1938) 199.
  %%CITATION = PRSLA,A165,199;%%

\bibitem{kal} T. Kaluza, Sitz. d. Preuss. Akad. d. Wiss. Physik-Mat. Klasse (1921),
966.

\bibitem{akk}P. G. O. Freund, Nuc. Phys. B. {\bf 209}, 146 (1982); K. Maeda, Class. Quant. Grav. {\bf 3}, 233 (1986);
 E. W. Kolb, M. J. Perry and T. P. Walker, Phys. Rev. D {\bf
33},869 (1986); P. Lor´e-Aguilar, E. Garc´i-Berro, J. Isern, and
Yu. A. Kubyshin, Class. Quant. Grav. {\bf 20}, 3885 (2003).

\bibitem{bd} C. H. Brans and R. H. Dicke, Phys. Rev. {\bf 124} (1961) 925.

\bibitem{gen} P. G. Bergmann, Int. J. Theor. Phys. {\bf 1} (1968), 25; R. V. Wagoner, Phys. Rev. D {\bf 1} (1970),
3209; K. Nordtvedt, Astrophys. J. {\bf 161} (1970), 1059.

\bibitem{bek} J. D. Bekenstein, Found. Phys. {\bf 16}, 409 (1986).

\bibitem{19}
 I. L.
Shapiro and J. Sola, JHEP 0202 (2002) 006; A. Babi\'c, B.
Guberina, R. Horvat, and H. \v{S}tefan\v{c}i\'c, Phys. Rev. D {\bf
65}, 085002 (2002);  I. L. Shapiro, J. Sola, C. Espana-Bonet, and
P. Ruiz-Lapuente, Phys. Lett. B {\bf 574}, 149 (2003); B.
Guberina, R. Horvat, and H. \v{S}tefan\v{c}i\'c, Phys. Rev. D {\bf
67}, 083001 (2003);
 C. Espana-Bonet, P. Ruiz-Lapuente, I. L. Shapiro and J.
Sola, JCAP {\bf 0402}, 006 (2004).


\bibitem{21} M. Reuter, Phys. Rev. D {\bf 57} (1998) 971;
A. Bonnano and M. Reuter, Phys. Rev. D {\bf65} 043508 (2002).

\bibitem{7}
  R.~Horvat,
  %``Holography and variable cosmological constant,''
  Phys.\ Rev.\  D {\bf 70}, 087301 (2004);
  %%CITATION = PHRVA,D70,087301;%%

\bibitem{Guberina}
B.~Guberina, R.~Horvat and H.~Nikolic, Phys. Rev. D {\bf 72},
125011 (2005).


\bibitem{Guberina0}
Y. Gong,  Phys. Rev. D, {\bf 70}, 064029 (2004);
  M.~R.~Setare, JCAP {\bf 0701}, 023, (2007).



%\cite{Zhang:2009ae}
\bibitem{Zhang:2009ae}
  Q.~J.~Zhang and Y.~L.~Wu,
  %``Dark Energy and Hubble Constant From the Latest SNe Ia, BAO and SGL,''
  arXiv:0905.1234 [astro-ph.CO].
  %%CITATION = ARXIV:0905.1234;%%


\bibitem{Horvat}
  B.~Guberina, R.~Horvat and H.~Stefancic,
  %``Hint for quintessence-like scalars from holographic dark energy,''
  JCAP {\bf 0505}, 001 (2005);
  %%CITATION = JCAPA,0505,001;%%
  B.~Guberina, R.~Horvat and H.~Nikolic,
  %``Dynamical dark energy with a constant vacuum energy density,''
  Phys.\ Lett.\  B {\bf 636}, 80 (2006);
  %%CITATION = PHLTA,B636,80;%%
  B.~Guberina, R.~Horvat and H.~Nikolic,
  %``Nonsaturated Holographic Dark Energy,''
  JCAP {\bf 0701}, 012 (2007).
  %%CITATION = JCAPA,0701,012;%%


%\cite{Galli:2009pr}
\bibitem{Galli:2009pr}
  S.~Galli, A.~Melchiorri, G.~F.~Smoot and O.~Zahn,
  %``From Cavendish to PLANCK: Constraining Newton's Gravitational Constant with
  %CMB Temperature and Polarization Anisotropy,''
  arXiv:0905.1808 [astro-ph.CO].
  %%CITATION = ARXIV:0905.1808;%%


%\cite{Xu:2009ss}
\bibitem{Xu:2009ss}
  L.~Xu and J.~Lu,
  %``Holographic Dark Energy in Brans-Dicke Theory,''
  Eur.\ Phys.\ J.\  C {\bf 60}, 135 (2009);
  %%CITATION = EPHJA,C60,135;%%
  L.~Xu, J.~Lu and W.~Li,
  %``Cosmic Constraints on Holographic Dark Energy in Brans-Dicke Theory,''
  arXiv:0905.4174 [astro-ph.CO].
  %%CITATION = ARXIV:0905.4174;%%





\end{thebibliography}
\end{document}